\documentclass[hidelinks]{article}
\usepackage{spconf,amsmath,graphicx}
\usepackage[english]{babel}

\usepackage{subcaption}

\usepackage{hyperref}
\hypersetup{
    colorlinks=false,
    linkcolor=blue,
    filecolor=magenta,      
    urlcolor=blue,    
    }

\usepackage[final]{microtype}

\usepackage{enumitem}
\setlist{nosep, leftmargin=14pt}

\renewcommand{\paragraph}[1]{\noindent\textbf{#1}\;}

\newcommand{\smallAddress}[1]{\vspace{-1.1mm}{\small #1}}

\def\sectionVspacePre{\vspace{-3mm}}
\def\sectionVspacePost{\vspace{-1mm}}

\def\subsectionVspacePre{\vspace{-2mm}}
\def\subsectionVspacePreExtra{\vspace{-2mm}}
\def\subsectionVspacePost{\vspace{-2mm}}

\title{Brain MRI Screening Tool with Federated Learning}
\name{%
\begin{tabular}{@{}c@{}}
    Roman Stoklasa$^{*,\dagger}$ \quad
    Ioannis Stathopoulos$^{*,\ddagger}$ \quad
    Efstratios Karavasilis$^{\|}$ \quad
    Efstathios Efstathopoulos$^{\ddagger}$\\
    Marek Dostál$^{\mathsection,\mathparagraph}$\qquad
    Miloš Keřkovský$^{\mathsection}$\qquad
    Michal Kozubek$^{\dagger}$\qquad 
    Luigi Serio$^{*}$%
    \thanks{This project was partially funded by the CERN budget for 
    Knowledge Transfer for the benefit of Medical Applications,
    by the Ministry of Health of the Czech Republic 
    (Grant No. NU21-08-00359) and the Ministry of Education, Youth and Sports of the Czech Republic (Project No. LM2023050).}
\vspace{-3mm}
\end{tabular}}

\address{%
  \smallAddress{$^{*}$ Technology Department, CERN, 1211 Geneva 23, Switzerland}\\
  \smallAddress{$^{\dagger}$ Centre for Biomedical Image Analysis, Faculty of Informatics, Masaryk University, Brno, Czech Republic}\\
  \smallAddress{$^{\ddagger}$ 2nd Dept of Radiology, Medical School, National \& Kapodistrian University of Athens, Chaidari, Greece}\\
  \smallAddress{$^{\|}$ Medical Physics Laboratory, School of Medicine, Democritus University of Thrace, Alexandroupolis, Greece}\\
  \smallAddress{$^{\mathsection}$ Dept of Radiology and Nuclear Medicine, Faculty of Medicine, Masaryk University and University Hospital Brno}\\
  \smallAddress{$^{\mathparagraph}$ Dept of Biophysics, Faculty of Medicine, Masaryk University, Brno, Czech Republic}
}

\begin{document}
%
\maketitle
%
%
\begin{abstract}
In clinical practice, we often see significant delays between MRI scans and 
the diagnosis made by radiologists, even for severe cases. 
In some cases, this may be caused by the lack of additional information and 
clues, so even the severe cases need to wait in the queue for diagnosis.   
This can be avoided if there is an automatic software tool, which would 
supplement additional information, alerting radiologists that 
the particular patient may be a severe case. 
%
We are presenting an automatic brain MRI Screening Tool and we are demonstrating 
its capabilities for detecting tumor-like pathologies. 
It is the first version on the path toward a robust multi-pathology 
screening solution. The tool supports Federated Learning, 
so multiple institutions may contribute to the model without disclosing 
their private data. 
\end{abstract}
%
%
%
%


\sectionVspacePre
\section{Introduction}%
\label{sec:intro}%
\sectionVspacePost

At clinical routine, radiologists typically deal with an overwhelming amount of MRI exams, 
often without prior knowledge of the case's context. In the evaluation queue, 
there are normal cases, abnormal, and often quite severe abnormal cases. 
Typically, patients must wait for a delivery of a verified diagnosis significant time. 
However, fast access to diagnosis means time-saving 
or even life-saving therapeutic or pharmaceutical decisions.

In clinical practice, one can easily find the lack of uniform protocol for 
examination prioritization. 
Radiologists can judge the priority of diagnosis by several factors, e.g., 
the clinical state of the patient, previous evaluated imaging exams, patient history, 
indications of the referral and the radiographer's initial findings during the acquisition.  
Unfortunately, these criteria are not always available, or they are barely applied 
under intense workload conditions, so it happens pretty often that radiologists 
take patients for diagnosis in chronological order. 

The goal of our work is to develop a Screening Tool, software that would automatically 
evaluate all brain MRI scans in a given 
hospital, 
and which would produce pre-diagnostic reports for radiologists.
Based on such reports, radiologists could easily decide which examinations need to 
be processed sooner and with higher priority, or, they might decide to 
process the ``easy cases'' first (i.e., cases that can be completed quickly and easily), 
to increase diagnostic throughput. 
The ultimate goal is to help decrease the waiting time between the scan and the diagnosis, 
especially for severe cases, by assisting radiologists to work more efficiently with better prioritization. 
The pre-diagnostic report predicts whether the patient's brain contains 
pathological regions, and if so, how large these regions are and where they are located. 
The report also presents important slices that support the tool's prediction. 

We used a commonly known machine learning (ML) approach with deep convolutional 
neural networks to develop the Screening Tool. 
To train ML models to a decent level of performance, a large number of training examples
are needed, accompanied by manual labels made by expert radiologists. 
The problem is that the creation of manual labels is costly in terms of 
required time and work done by experts in the field. 
Also, if we want to develop ML models that would work for a wide range of clinics in practice, 
we would ideally need examples from many institutions. 
This is often very problematic since medical data are personal and strictly confidential. 
Thus, we used Federated Learning to overcome both problems.

Federated Learning (FL) \cite{konecny2015federated, first_fl_article} is a technology that enables
the distributed training of ML models with a set of remote devices (nodes) without 
sharing or disclosing the training data. The data remains securely stored within 
individual nodes (e.g., hospitals), 
and only the training updates and 
models are shared among nodes (inside the federation). 
FL has been successfully demonstrated multiple times for brain MRI images \cite{bernardoFL2022, fets2022}. 


Thanks to the FL approach, multiple institutions can team up and form a federation where each institution can supply fewer samples. 
The manual labels for a limited number of samples can be produced more easily by radiologists at those institutions, 
and since they can be prepared at many institutions in parallel, the overall time until a suitable model can be trained is much less compared 
to the situation where each institution would like to develop its own model. 
Together with data privacy, this is one of the most significant advantages of the FL approach for practical applications.

This paper presents our developed Screening Tool for brain MRI images, trained with Federated Learning, 
demonstrating impressive results. Thanks to the autonomous and fully automatic processing, this tool 
can be deployed in multiple institutions to process every brain MRI scan, producing clear 
pre-diagnostic reports for radiologists. Generated screening results can be used to draw radiologists' 
attention to cases that need to be diagnosed sooner and/or require more focus. 
While the tool internally performs brain anomaly segmentation, the default presentation 
for users is in the form of detections (bounding boxes) for the purpose of enhanced clarity. 



\sectionVspacePre%
\section{Materials and Methods}%
\label{sec:methods}%
\sectionVspacePost%

\subsectionVspacePre
\subsection{Screening Tool}
\subsectionVspacePost
Our aim is to develop a Screening Tool software, whose responsibility would be to 
automatically process every brain scan from an MRI machine in a clinic and produce 
screening report for radiologists. The reports should indicate if there are 
any pathologies inside the patient's brain, and if yes, to indicate their location, 
size and visual appearance. 
The software is intended to operate independently and to process the scans either 
immediately after acquisition or in batches, e.g., over the night, to prepare a set 
of reports before radiologists start their new working day. 

The heart of the tool is the deep learning convolutional neural network. 
For the currently presented version, 
we decided to use U-Net \cite{ronneberger2015unet} with Inception-v3 
\cite{inceptionv3} as a backbone, utilizing pre-trained encoder weights. 
The implementation can be found in the Segmentation Models library \cite{segmentation-models-git}.
The model processes individual 2D axial slices separately. 
We selected this smaller and simpler model for 2D processing to  
decrease the computational requirements for the individual FL training nodes. 
Models that process 3D volumes (e.g., nnU-Net \cite{isensee2021nnu}) often have very 
high GPU memory requirements, which may not be available in some medical institutions.

The whole data processing pipeline is composed of multiple steps. 
First, required MRI modalities are selected from the MRI scan package. 
We designed our current model to work with three modalities: FLAIR, T2, and T1ce. 
T1ce can be substituted by standard T1 if the scan with contrast was not 
performed for a particular patient. 
Next, automatic skull-stripping and co-registration of all modalities are performed. 
The registration produces only the transformation matrices that are used later 
for resampling.
Next, the intensities of all modalities are normalized using the method that fits the 
cumulative distribution function (CDF) of image intensities to a pre-computed template. 
This approach is similar to histogram matching \cite{gonzalez2008DIP}, but we allow 
only the linear scaling and uniform shifting of the intensity values. 
This restricted histogram matching method preserves local relationships of intensities 
inside one image, while the overall distribution of intensities is very similar across individual 
examinations. Each modality is normalized separately, and the 
desired templates are derived from the training set.  
Then, 2D axial slices are extracted from the modality with the highest axial resolution,
and are resampled to a common desired pixel resolution (0.72 mm/px in our case).
This normalizes depicted brain size across different examinations and different institutions. 
Finally, the DL model predicts pathologies for every 2D slice independently by producing 
the binary mask of detected pathology.
The last step is to combine all of these results together and to produce the report. 
Then, the report is delivered to radiologists by a suitable method, e.g., either be saved to 
an appropriate location, it can be, e.g., sent by email or delivered by any other means. 

\begin{figure*}[t]
    \centering
    \includegraphics[height=87mm,width=0.95\textwidth]{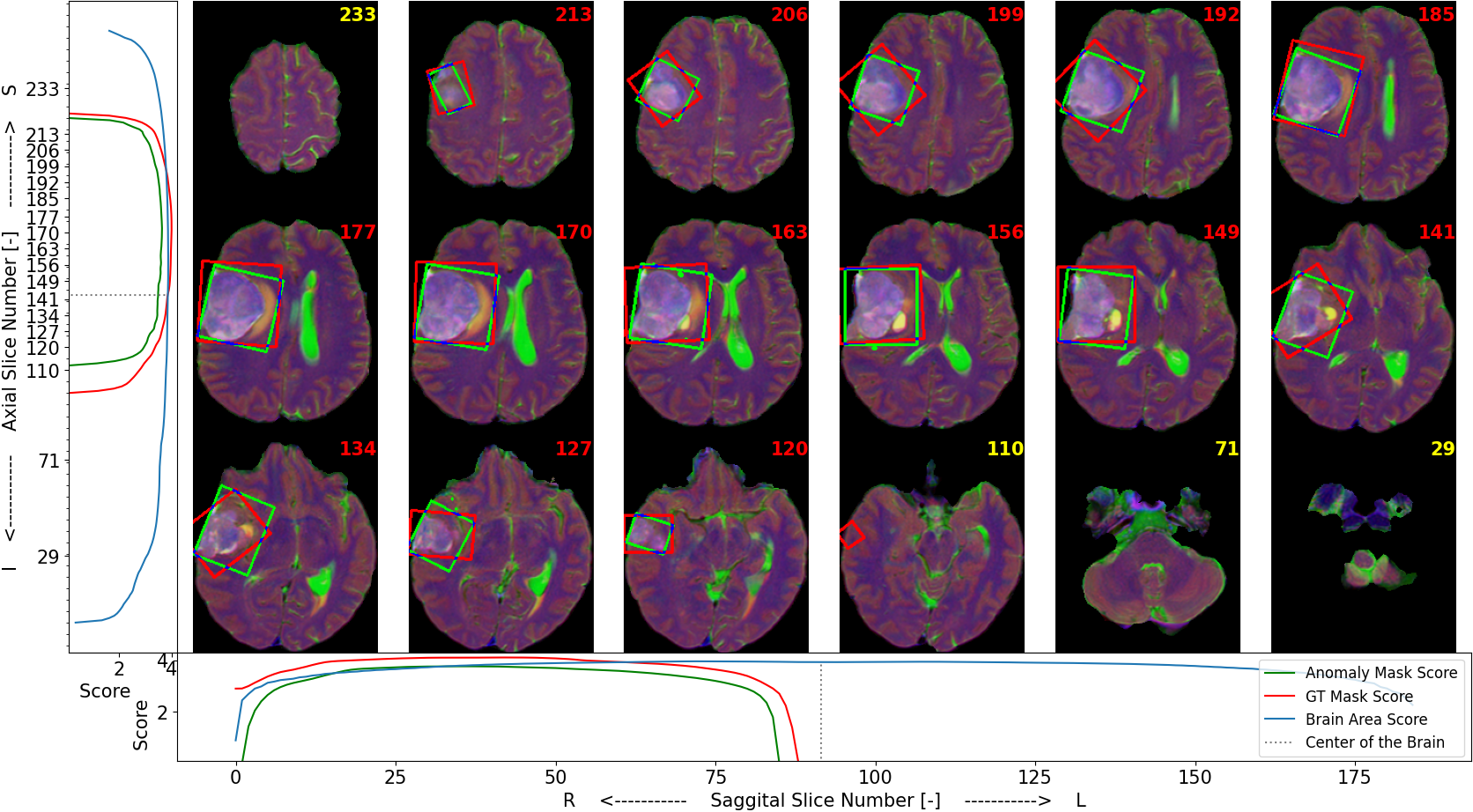}
    \caption{ 
    Example of output from our Screening Tool. The radiologist is presented with abnormality scores along the axial and sagittal axis and the 18 most interesting slices that should support model findings. 
    Numbers displayed along the vertical Y axis correspond to depicted slides in the main display area, and the value represents the index of the axial slide. The GREEN boxes show what the Screening Tool predicted, 
    while REDs are derived from manual labels.
    }
    \label{fig:screening-result}
\end{figure*}

The report, as depicted in Fig~\ref{fig:screening-result}, contains three main information areas: 
scores of detected pathologies along the axial and the sagittal axis,
and the main area, where the 18 most ``interesting'' axial slices are depicted. 
The slides are displayed as RGB color images, mapping intensities from FLAIR, 
T2 and T1ce modalities to red, green, and blue color channels, respectively. 
The tool tries to select the most representative mix of abnormal and normal slices. 
If some pathologies are detected, slices containing them are prioritized 
for display, but the tool also tries to display some surrounding normal ones 
for a better perception of the context. 
Detected anomalies are highlighted by oriented bounding boxes, which are placed 
around the detected object to avoid occluding the anomaly and its boundaries. 
If the user prefers, the tool can also display precise segmentation boundaries
obtained from the U-Net model, as well as display sagittal or coronal preview slices
(not shown here).

\subsectionVspacePre\subsectionVspacePreExtra%
\subsection{Federated Learning}%
\subsectionVspacePost%
One of the key features of our solution is the support of Federated Learning. 
We build the Screening Tool on top of CAFEIN FL Platform \cite{cafein-fl-platform}, 
developed at CERN. This platform is implemented with a focus on practical 
applicability and usability in real-world situations, and is easily adaptable to any 
domain or application. 

The communication between the nodes is implemented through the MQTT protocol
(Message Queuing Telemetry Transport). 
In this use-case, we chose the topology with Parameter Server, though other 
fully-distributed and decentralized topologies are also possible \cite{bernardoFL2022} 
in the future. 

The CAFEIN FL platform supports proper real-world network communication. 
The federation, formed to train and evaluate the experiments described below, 
was set up on multiple distinct physical computers distributed across Europe.

\subsectionVspacePre\subsectionVspacePreExtra%
\subsection{Data}%
\subsectionVspacePost%
For the development and demonstration of the Screening Tool, we used real clinical data 
from two hospitals in two different countries: 
Aiginiteio University Hospital in Athens, Greece, and University Hospital Brno, Czech Republic. 
All examinations contain three modalities: FLAIR, T2, and T1ce (contrast-enhanced T1).

We decided to start with examinations that contained tu\-mor-like anomalies 
for this first version of the Screening Tool.  
We included various types of tumors in the dataset, 
namely High-Grade Glioma (HGG), Low-Grade Glioma (LGG), Lymfoma, Meningeoma, and metastatic tumors. 
This is a significant difference between our dataset and other public datasets, 
e.g. BraTS, which contains only HGG and LGG. 
In total, our dataset contains 160 examinations, where most of the exams 
were used for testing, and only the smaller part for training 
(more details in Section \ref{sec:results}).


\sectionVspacePre%
\section{Results and Discussion}%
\label{sec:results}
\sectionVspacePost%


\subsectionVspacePre%
\subsection{Experiment setup}%
\subsectionVspacePost%
The experiment we conducted is a demonstration of how a Federated Learning 
can boost the gains for institutions if they decide to join the federation
despite the fact that each institution contributes only a limited number of 
training data. 

In our setup, we devoted only 58 MRI examinations to the training dataset, 
while the remaining 102 exams were left for validation and testing. 
This division was chosen on purpose to demonstrate the vitality of FL training, 
where each node (hospital) has just a limited number of annotated examples.

The 58 training exams were divided into 8 clients, each possessing 
between 6 and 8 examples. The data from Athens and Brno were not 
mixed together, so there were 4 FL clients backed up with Athens data, 
and 4 clients with Brno data. 
Since the clients had only a limited number of examinations, an extensive data 
augmentation was used (affine and elastic transforms, flipping). 
The augmentation ratio was set to 1:19, i.e., for every 
original axial slice taken for the training, 19 transformed slices were generated. 

The Parameter Server was responsible for orchestrating the training, 
and for merging model updates from individual clients.
We used Federated Averaging \cite{first_fl_article} in this experiment. 
The server was also allowed to throttle the fast clients 
in order to balance the overall number of ``samples seen'' 
(i.e., the number of training inputs) between individual clients.

The model was trained during 200 FL rounds, while each round was set to last approx. 6 minutes. 
The learning rate was progressively decreasing during the training.  
In total, the final model was trained with approx. 13 mil. samples (slices) seen 
(i.e., the sum of samples seen from all 8 nodes). 

The model was then tested on 102 examinations. We are reporting on two main metrics:
i) the Exam-averaged Dice coefficient (EaD), and ii) the Global Dice (GD) coefficient. 
The Exam-averaged Dice is derived as the average of Dice coefficients computed for 
each MRI exam separately. This metric stresses out errors that occur inside 
examinations with small anomalies, as these would inflict a higher penalty on the final score.  
To counteract this bias, we report also 
the Global Dice coefficient, which is computed over the whole volume of all 
examinations together (i.e., as if all the examinations would be concatenated 
into one huge 3D image). 
The complete evaluation of a single MRI examination took approx. 9 seconds using 
a rather old GTX 1080 GPU and Intel i7-8700K CPU.

To understand better our achieved performance in a context, we compare the results 
of our FL model with two alternative scenarios: i) the node would like to train its 
own model using only its own limited data, and ii) the node would like to use 
some available ``off-the-shelf'' model.


\begin{table}[t]
    \centering
\caption{Validation results obtained on 102 MRI exams.}
\label{tab:results}
    \begin{tabular}{|c||c|c|} \hline 
 & Exam-avrg Dice&Glob. Dice\\ \hline\hline
         FL clients (average)&  \textbf{0.837}& \textbf{0.884}\\ \hline 
 Single client (max)& 0.744&0.817\\ \hline 
         Rixez (BraTS) \cite{rixezBrats2021} &  0.816& 0.883\\ \hline
    \end{tabular}
\end{table}

\subsectionVspacePre\subsectionVspacePreExtra%
\subsection{Results}
\subsectionVspacePost%
We evaluated our model on the complete set of 102 examinations, simulating that 
these exams are the new scans at one of the nodes. 
Since each node stores its locally-best model, the models across nodes may differ slightly. 
Thus, we performed the evaluation on each of the 8 clients and took the average. 
The Screening Tool trained using Federated Learning and all 8 clients 
achieves EaD score of 0.84 and GD score of 0.89, see Table~\ref{tab:results}. 

We also evaluated the situation when each node would train 
its own private model using only its small dataset. 
The training protocol and parameters 
were the same as in the FL training. 
We tested each of the 8 models against the testing set, 
but the best-achieved Dice scores are significantly lower than in the FL case: 
reaching 0.74 for EaD and 0.82 for GD. 

For the experiment with the off-the-shelf model, we took the winning submission 
to the 2021 Brain Tumor Segmentation Challenge \cite{rixezBrats2021,rixez-github},
which we denote as ``Rixez model''. 
We ran the prediction of the model and performed the same evaluation. 
The achieved results of the Rixez model are slightly worse than ours for EaD, 
and on par with ours for the GD metric. 
It is also worth noting, that the processing time of the Rixez model 
was significantly higher, with approx. 45 seconds per exam 
on a much faster NVIDIA A100 GPU.

\subsectionVspacePre\subsectionVspacePreExtra%
\subsection{Discussion}%
\subsectionVspacePost%
The results show that Federated Learning may be a very useful concept, how 
different institutions may obtain useful machine learning model/tool, requiring
only a limited amount of annotated data contributed to the federation. 
Since the amount of data required from any node is small, the whole federation 
can benefit from a trained model in much less time compared to the situation, 
when the institution would like to train its own model. 
Federated Learning ensures the privacy of the data, so the institutions need not 
to worry about collaborating with many partners. 

Also, our results show that training a custom model still makes sense. 
The off-the-shelf BraTS model, trained to a slightly different task, 
works almost equally well for the tumor-like pathologies in our current dataset. 
However, the off-the-shelf model might become a serious limitation in the near 
future when we plan to extend the screening tool for various pathologies 
(e.g., white matter hyper-intensity, multiple sclerosis, and possibly also brain strokes).

As shown in Fig~\ref{fig:results-individual}, our model works well for various tumor types. 
However, as depicted in Fig~\ref{fig:result-samples-WMH}, the model sometimes detects also 
some other brain pathologies. 
This is the correct output for the practical use of the Screening Tool, 
but currently, it decreases our Dice scores since these types of pathologies are 
not marked by manual labels. 

\def\figTwoCellWidth{28.5mm}

\begin{figure}[t]%
     \centering%
     \begin{subfigure}[b]{\figTwoCellWidth}
        \centering%
        \includegraphics[height=3.4cm,width=28mm]{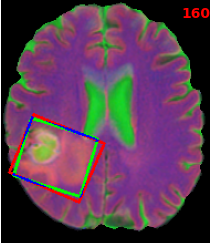}
        \caption{HGG}
        \label{fig:result-samples-hgg}
     \end{subfigure}%
     \begin{subfigure}[b]{\figTwoCellWidth}
        \centering
        \includegraphics[height=3.4cm]{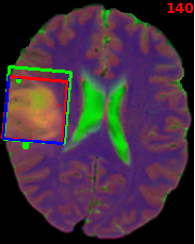}
        \caption{LGG}
        \label{fig:result-samples-lgg}
     \end{subfigure}%
     \begin{subfigure}[b]{\figTwoCellWidth}
         \centering
         \includegraphics[height=3.4cm,width=2.85cm]{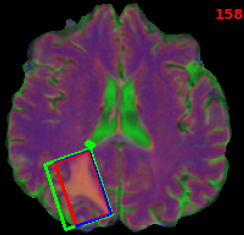}
        \caption{Lymfoma}
        \label{fig:result-samples-lymf}
     \end{subfigure}%
     \\%
     \begin{subfigure}[b]{\figTwoCellWidth}
         \centering
         \includegraphics[height=3.3cm,width=28mm]{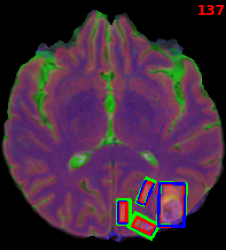}%
        \caption{Metastatic}
        \label{fig:result-samples-meta}
     \end{subfigure}%
     \begin{subfigure}[b]{\figTwoCellWidth}
         \centering
         \includegraphics[height=3.3cm,width=28mm]{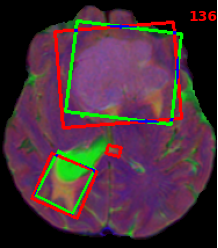}
        \caption{Meningeoma}
        \label{fig:result-samples-meng}
     \end{subfigure}%
     \begin{subfigure}[b]{\figTwoCellWidth}
         \centering
         \includegraphics[height=3.3cm,width=28mm]{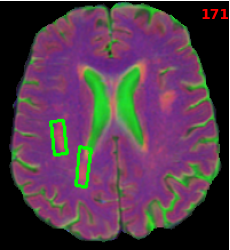}
        \caption{WMH}
        \label{fig:result-samples-WMH}
     \end{subfigure}%
    \caption{
    Examples of detections for various tumor types. GREEN rectangles represent Screening Tool prediction, 
    REDs are derived from manual labels, and BLUE color is where they overlap. 
    As shown in Fig.~\ref{fig:result-samples-WMH}, the model sometimes 
    de\-te\-cts also other brain pathologies, here White matter hyperintensities (WMH), 
    which were not manually labeled but are considered as correct detections 
    from the radiological perspective. 
    }
    \label{fig:results-individual}
\end{figure}


\sectionVspacePre%
\section{Conclusion}%
\label{sec:conclusion}%
\sectionVspacePost%

We presented our automated brain MRI Screening Tool, 
tra\-ined with federated learning with 8 clients, 
which has been tested and validated for tumor-like pathologies. 
The tool exhibits very good sensitivity and specificity, which gives
promises for practical usefulness for practical diagnosis prioritization 
once the tool is deployed for real clinical use. 

In the near future, we are aiming to enrich the model to be capable of 
detecting reliably multiple brain pathologies (e.g., white matter 
hyperintensities, multiple sclerosis, brain strokes, and others). 
Also, we plan to evaluate the effect of this tool on diagnosis prioritization 
in real clinical practice and its impacts on medical care improvement.

\bibliographystyle{IEEEbib}
\bibliography{references}

\sectionVspacePre%
\section{Compliance with ethical standards}%
\label{sec:ethics}%
\sectionVspacePost%

This study was performed in line with the principles of the Declaration of Helsinki. 
Approval was granted by the Ethics Committee of University Hospital Brno.

\end{document}